\documentclass[12pt,twocolumn]{article}
\usepackage[T1]{fontenc}
\usepackage[utf8]{inputenc}
\usepackage[english]{babel}
\usepackage{float}
\usepackage{color}
\usepackage[caption = false]{subfig}
\usepackage{amsfonts,amsmath,amssymb}
\usepackage{graphicx}

\begin{document} 
\onecolumn

\newcommand{\Jaa}{J_{AA}}
\newcommand{\Jbb}{J_{BB}}
\newcommand{\Jab}{J_{AB}}
\newcommand{\Tobd}{T_{\rm ObD}}
\newcommand{\Jeff}{J_{\rm eff}}
\newcommand{\Heff}{H_{\rm eff}}
\newcommand{\comment}[1]{}
\renewcommand{\contentsname}{Table of contents}
\newcommand{\comments}[1]{}   

\title{Methods for detecting Order-by-Disorder transitions: \\
the example of the  Domino model
}
\author{Hugo Bacry,$^1$ Leticia F. Cugliandolo$^{1,3}$ and Marco Tarzia$^{2,3}$ \\ 
  {\normalsize $^1$Sorbonne Université, Laboratoire de Physique Théorique et Hautes Energies,}\\
    {\normalsize CNRS  UMR  7589, 4, Place Jussieu, Tour 13, 5\`eme étage, 75252 Paris Cedex 05, France}\\
   {\normalsize $^2$Sorbonne Université, Laboratoire de Physique Théorique de la Matière Condens\'ee,}\\
   {\normalsize CNRS  UMR  7600, 4, Place Jussieu, Tour 13, 5\`eme étage, 75252 Paris Cedex 05, France}\\
  {\normalsize $^3$Institut Universitaire de France, 1, rue Descartes, 75231 Paris Cedex 05, France}
  }
\maketitle

\begin{abstract}
Detecting the zero-temperature thermal Order-by-Disorder transition in classical magnetic systems is notably difficult. 
We propose a method to probe this transition in an indirect way.
The idea is to apply adequate magnetic fields to transform the zero temperature transition into a finite temperature 
sharp crossover, which should be much easier to observe and characterise with usual laboratory methods.
\end{abstract}

\newpage

\tableofcontents

\newpage

\section{Introduction}

In condensed matter physics, fluctuations, whether thermal or quantum, usually 
suppress order. However, this is not a rigorous rule. Some systems undergo an 
``Order-by-Disorder''  (ObD) transition in which the fluctuations restore order in an otherwise 
disordered ground state~\cite{villain1980order,shender1996order}. 
This ObD transition is, more precisely, the mechanism whereby a system with a non-trivially degenerate ground state develops long-range order by the effect of classical or quantum fluctuations. Therefore, a classical system exhibiting this transition has no long-range order when the temperature is strictly zero and develops some at non-vanishing temperature.

This phenomenon was first exhibited in the classical 2D Domino Model~\cite{andre1979frustration}.   
An experimental 3D realisation, in the form of Ising pyrochlores with staggered antiferromagnetic 
order frustrated by an applied magnetic field was recently proposed~\cite{guruciaga2016field,guruciaga2019monte}
(concrete examples could be Nd$_2$Hf$_2$O$_7$ or Nd$_2$Zr$_2$O$_7$). 
Indeed, 
the zero temperature ObD transition is relatively common in highly frustrated magnetic models~\cite{diep2016theoretical,Chalker11}.
In this context, the geometry of the lattice and/or the nature of the interactions make the simultaneous 
minimisation of each term contributing to the energy impossible~\cite{moessner2006geometrical}. 
Two consequences of frustration are the increase of the ground state energy compared to the one of the 
unfrustrated model and the scaling of the number of degenerate ground states  (sublinearly) 
with the size of the system.


Although the reason for the classical ObD transition is clear, it has been very difficult to exhibit experimental evidence 
for it. One of the reasons is that the transition occurs at zero temperature 
and it is therefore difficult to establish whether order is selected through the ObD mechanism or it is due to energetic contributions 
not taken into account that actually lift the ground state degeneracy.
The aim of this paper is to propose a way to probe the ObD transition in 
an indirect way which should be relatively easy to implement in the lab. The idea, as we explain in the main 
part of the article, is to use external magnetic fields to transform the zero temperature transition into a 
finite temperature sharp crossover, or maybe even a genuine phase transition, and then detect the latter with usual methods.
For concreteness, we explain how this is achieved in the context of the 2D Domino Model.

The paper is organized as follows. In Sec.~\ref{sec:domino} we recall the definition and main properties of the 
Domino Model. In particular, we establish the effective 1D model that describes its low energy properties \cite{villain1980order}, 
which we will use in the rest of our study. 
In Sec.~\ref{sec:random} we add quenched disorder in the form of columnar random magnetic fields as a first attempt to displace the ObD transition 
to a finite temperature. We start by showing, with an Imry-Ma argument~\cite{imry1975random}, 
that such a  2D disordered model cannot have a finite 
temperature phase transition but just a crossover. Still, we characterise the pseudo ferromagnetic order
 thus achieved studying a random field 1D effective model with the renormalization group approach.
The next strategy, described in Sec.~\ref{sec:staggered}, is to use alternate columnar magnetic fields. With them 
we achieve the goal of finding a finite critical temperature  but we lose a bit of the phenomenology of the ObD transition,
as we explain in the body of the paper. In each Section we analyse the quench 
dynamics of the pure and disordered Domino Models using Monte Carlo simulations, and we describe how the 
temporal evolution confirms the static behaviour expected asymptotically. A Section with our conclusions closes the 
article.

\section{The Domino Model}
\label{sec:domino}

The Domino Model is a 2D model defined on a square lattice with two kinds of ions 
A and B that carry Ising spins and are placed on alternating columns~\cite{villain1980order,andre1979frustration}. 
There are thus three different interactions $\Jaa$, $J_{BB}$ and $ J_{AB}$ between nearest neighbor spins. $\Jaa$ and $ J_{AB}$ are ferromagnetic ($\Jaa>0$, $\Jab >0$) while $J_{BB}$ is antiferromagnetic ($\Jbb<0$). With these parameters, all plaquettes in the lattice are  frustrated. The system has
 size $N\times N$ ($N/2$ columns A and $N/2$ columns B each of length $N$) and we assume periodic boundary conditions.
Therefore, the Hamiltonian is
\begin{equation}
H = \Jab\sum_{i,j} s_{i,j} s_{i,j+1} + \Jaa\sum_{\substack{i\\j\,even}} s_{i,j} s_{i+1,j} +  \Jbb\sum_{\substack{i\\j\,odd}} s_{i,j} s_{i+1,j}
\; , 
\end{equation}
with $s_{i,j} = \pm 1$ the Ising spins sitting on the vertices of the square lattice. 
Henceforth, the rows are
labeled  $i=1,2,...,N$ and the columns are labeled $j=1,2,...,N$.
We assume that the interactions respect the hierarchy
\begin{equation}
\Jaa\gg  \lvert J_{BB} \rvert > J_{AB}  
\; .
\label{eq:hierarchy}
\end{equation}

\subsection{Ground and first excited states}

With the choice of parameters in Eq.~(\ref{eq:hierarchy}), the ground states have ferromagnetic order on each A column and antiferromagnetic order on each B column. Moreover, A and B columns 
are effectively uncoupled because half of the spins are up and half down in a B 
column. As a consequence, it has the same cost for the A columns to be up or down. The ground states are frustrated because only half of the horizontal bonds with coupling constant $\Jab$ can be satisfied in an 
optimal configuration. Moreover, looking at a typical ground state as the one displayed in Fig.~1(a), we can see that the A columns are either up or down and the spins on the B columns alternate 
between up and down, yielding a vanishing global magnetization $M = 0$ (in the $N\to\infty$ limit).
The same occurs in all $T=0$ ground states. It is easy to see that there are $2^N$ such ground states.  
The ground state entropy is then sub-extensive, $S\propto N$, but still much larger than the usual ${\mathcal O}(1)$ one of, 
say, the 2D ferromagnetic Ising model.

\begin{figure}[h!]
\hspace{2.5cm}
(a) \hspace{5cm} (b)
\begin{center}
\includegraphics[scale=0.45]{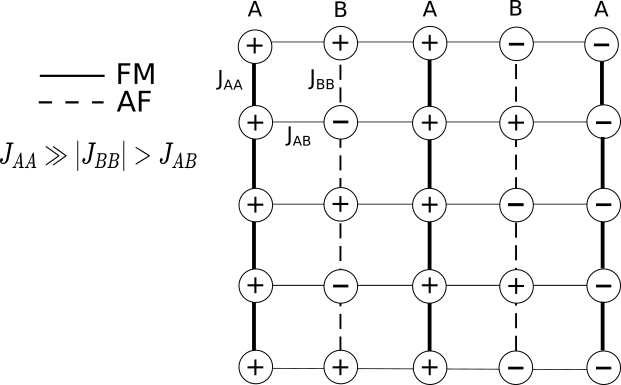}
\hspace{0.5cm}
\includegraphics[scale=0.45]{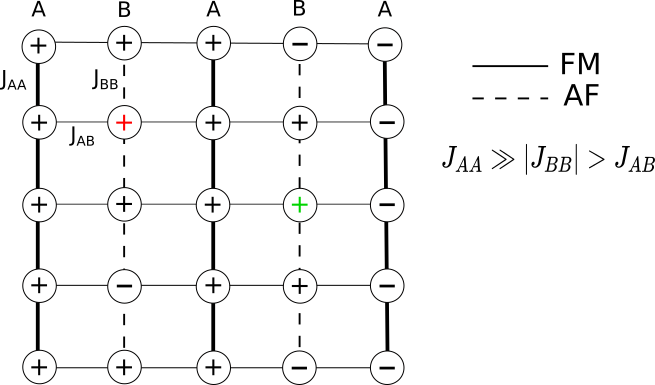}
\caption{\small 
(a) A typical ground state of the Domino Model. Ferromagnetic interactions are represented by full lines, antiferromagnetic interactions by dashed lines. The hierarchy in Eq.~(\ref{eq:hierarchy}) is illustrated with bold and thin lines.
(b) Two possible excitations are highlighted in red and green. The red one has lower energy than the green one because it is sandwiched between two A columns of the same sign. Flipping a spin of the A columns would cost even more energy because of the hierarchy in the coupling constants in Eq.~(\ref{eq:hierarchy}).
}
\end{center}
\end{figure}

Starting from the ground state, we can construct the first excited state by taking a B column sandwiched in between two A columns of the same sign and turning one of its spin from being anti-aligned to being aligned with the A spins, 
see the red + in Fig.~1(b). We see that we lose $4\lvert J_{BB} \rvert$ energy and we gain $4J_{AB}$ energy from this process. 
Fixing $E_{\rm GS}=0$ for the ground state energy, the excited state has energy $E=\epsilon_1 = 4(\lvert J_{BB} \rvert - J_{AB})$. The other possible excitation in a B column is one in which the flipped spin is in between two anti-aligned A columns,
see the green + in Fig.~1(b), which has energy $\epsilon_2 = 4 |J_{BB}|$ and it is a higher excited state than the previous one. 
So, at low (but finite) temperature, when only the first excited states are statistically relevant, A columns tend to be aligned for these excitations to exist. This entropic effect forces the system to have long range ferromagnetic 
order of the A columns at low temperature and thus exhibit the zero temperature ObD transition~\cite{villain1980order}. Order is maintained
until the critical temperature $T_c=1/\beta_c$ (we set $k_B=1$) given by
\begin{equation}
\sinh(2\beta_c J_{AB}) \sinh (\beta_c |J_{AA}+J_{BB}|) = 1
\label{eq:critical-temp}
\end{equation}
beyond which the system becomes a conventional paramagnet.


\subsection{The effective 1D model}
\label{sec:eff_1D}

An effective 1D model for the low temperature, $T\ll \epsilon_2$, properties of the system that focuses on the  
A columns was derived in~\cite{villain1980order}. The argument goes as 
follows. First, since $\Jaa$ is much stronger than the two other couplings, 
see Eq.~(\ref{eq:hierarchy}), one assumes that the A columns are perfectly aligned and then represents them 
as macro-spins of value $N$. 
Second, \\
-- If a B chain is sandwiched in between two A chains with parallel spins, the first excitations have energy 
\begin{equation}
\epsilon_1 = 4(\lvert J_{BB} \rvert - J_{AB})
\; , 
\end{equation} 
and $N/2$ of them 
are possible, as explained in the previous Subsection. 
The partition function of the 
B chain in this background (that we indicate with the subscript $F$) is
$$
Z_F \simeq [1 + \exp(-\beta \epsilon_1)]^{N/2}
\; .
$$
-- If, instead,  the two A chains have opposite orientation, the second $N/2$ excitations have energy $\epsilon_2 = 4\lvert \Jbb \rvert$. In this 
other background (that we label $AF$) the partition function is
$$Z_{AF} \simeq [1 + \exp(-\beta \epsilon_2)]^{N/2} \; . $$ 

We can now integrate out the spins of the B columns to get an effective nearest-neighbor 
coupling $\Jeff$ between the spins of two nearby 
A chains.
Thinking in terms of a 1D effective model of size $N/2$, the probabilities $P_F$ of two neighboring 
A chains (of size $N$) being parallel, and $P_{AF}$ of two neighboring  A chains being anti-parallel, are
$$
P_F = \frac{\exp(\beta\Jeff)}{2\cosh(\beta \Jeff)} \qquad \text{and} \qquad P_{AF} 
= \frac{\exp(-\beta \Jeff)}{2\cosh(\beta \Jeff)}
\; ,
$$
respectively.
On the other hand, the same probabilities in the original model are
$$
P_F = \frac{Z_F}{Z_F + Z_{AF}} \qquad \text{and} \qquad P_{AF} = \frac{Z_{AF}}{Z_F + Z_{AF}}
\; .
$$
Using these equations we find that
$$
\frac{P_F}{P_{AF}} = \exp(2\beta \Jeff) = \frac{(1 + \exp(-\beta \epsilon_1))^{N/2}}{(1 + \exp(-\beta \epsilon_2))^{N/2}} \simeq (1 + \exp(-\beta \epsilon_1))^{N/2}
$$
since we choose $\lvert \Jbb \rvert$ of the same order as $\Jab$ which makes  $\epsilon_1 = 4(\lvert J_{BB} \rvert - J_{AB}) \ll \epsilon_2 = 4\lvert J_{BB} \rvert$.
In conclusion we find a temperature dependent and ${\mathcal O}(N)$ effective coupling constant
\begin{equation}
\Jeff(\beta,N) = \frac{N}{4\beta}\ln[1+ \exp(-\beta \epsilon_1)]
\label{eq:Jeff}
\end{equation}
and the effective Hamiltonian of the 1D system is
\begin{equation}
\Heff(T) = -\Jeff(T,N) \sum_{j=1}^{N/2} s_j s_{j+1}
\label{eq:Heff1D}
\end{equation}
with the new ${\mathcal O}(1)$ Ising spins, $s_j=\pm 1$, representing the $N/2$ A columns.

We see that although the model in Eq.~(\ref{eq:Heff1D}) is one dimensional, the coupling constant is of 
macroscopic order ($\propto N$), allowing for long-range order in the effective model
that represents the ordering of the 2D system.
In this way, as soon as $T>0$, $\Jeff >0$ forcing the system  into a ferromagnetic phase as a regular 
2D ferromagnetic Ising model, even though only the A columns are ferromagnetically ordered: in the thermodynamic limit, 
the global magnetisation density $m=N^{-1}\sum_{j=1}^{N/2} s_j$ jumps from $0$ to $1/2$
in a discontinuous way. 
This approximation is valid as long as we use the hierarchy of coupling constants in 
Eq.~(\ref{eq:hierarchy}) and $\lvert \Jbb \rvert \sim \Jab$. Indeed, we need $\Jaa \gg (\lvert \Jbb\rvert, \Jab)$ to consider the A columns as macro-spins and $\epsilon_1 = 4(\lvert J_{BB} \rvert - J_{AB}) \ll \epsilon_2 = 4\lvert J_{BB} \rvert$ to keep only the first excitation accessible at the temperatures we study.

\subsection{Dynamics}

As far as we know, the stochastic evolution of the  kinetic 2D Domino Model has not been studied in detail yet. We will do it in
later Sections of this paper, where we will compare it to the ones of the disordered models.

\section{Columnar random fields}
\label{sec:random}

Let us add quenched disorder to the 2D Domino Model in the form of $N/2$ columnar random magnetic fields $h_{i,j}$
that couple bilinearly to the spins, $\sum_{i,j} h_{i,j} s_{i,j}$, but only to those on the A columns
and independently of the row index. In order words, $h_{i,j} = h_j \neq 0$ 
only for $j$ even. The $h_j$'s are random i.i.d. variables drawn from a Gaussian distribution $\mathcal{N}(0,\sigma^2)$. The typical local random fields take absolute values of the order of $\sigma = {\mathcal O}(1)$. 

The Imry-Ma argument can be easily applied to show that such disordered 2D model cannot have a phase transition, as we 
discuss below (Sec.~\ref{sec:ImryMa}). 
Nevertheless, the finite size model can still present a finite temperature crossover from a disordered low temperature 
state to a quasi ferromagnetically ordered state 
 at a higher temperature, in a way that mimics the ObD transition but at a non-zero temperature, before disordering it
again at a still higher temperature. We analyse the first 
crossover in the context of the effective 1D model  that we assume remains the same as the one 
derived in Sec.~\ref{sec:eff_1D}, represented by the $\Jeff$'s in Eq.~(\ref{eq:Jeff}),
even under the random fields which are, therefore, supposed to be very weak compared to $J_{\rm eff}$
(Sec.~\ref{subsec:1Ddisordered}). 
Finally, we study the quench dynamics of the 2D model using different initial states and final temperatures mostly in the 
region with quasi ferromagnetic order (Sec.~\ref{subsec:dynamics}).

\subsection{The Imry-Ma argument}
\label{sec:ImryMa}

Here we show that by extending the   
Imry-Ma argument~\cite{imry1975random} to this model, the ferromagnetic phase of the 
2D system should be destroyed by the addition of the columnar
random magnetic fields.
\comments{
The Imry-Ma argument states that, for a usual RFIM in dimension $d$ with Hamiltonian 
$$H = -\sum_{\langle i,j \rangle}Js_i s_j - \sum_{i}h_is_i$$
with $\langle i,j\rangle$ designating the nearest-neighbors and $h_i$ i.i.d. random variables, with $\mathbb{E}[h_i] = 0$, and $\mathbb{E}[h_i^2] = \sigma^2$, if we flip a spin domain $\Omega$ of characteristic length $l$ starting from an ordered phase we loose: $\epsilon_{boundary} \simeq 2J\partial\Omega$ but we can gain energy from the bulk: $\epsilon_{bulk} \simeq -2\sum_{i \in \Omega} h_i$ giving a total variation of energy
\begin{equation}
\Delta E \simeq 2J\partial\Omega  - 2\sum_{i \in \Omega} h_i 
  \sim 2Jl^{d-1} - 2l^{d/2} Y
\end{equation}
with $\partial\Omega$ designating the border of the $\Omega$ domain and $Y = \mathcal{N}(0,\sigma^2)$ making use of the CLT in the second line. Comparing both terms we see that for $d>2$, there is no macroscopic domain of size $l$ that we could flip without increasing the energy of the system ($\Delta E$ is positive for large $l$). Instead for $d=2$ (and lower), we can have $\Delta E < 0$ so we can lower the energy by flipping a macroscopic domain of spins. Therefore, no ordered phase can exist in the thermodynamic limit. This argument estimates the lower critical dimension to $d=2$ for the RFIM.
}

Let us sketch why this is so. In order to simplify the discussion, take a homogeneous 
ferromagnetically coupled ($J_{AB}=J_{AA}=J_{BB}=J>0$)  Ising model in 2D with 
columnar random fields. The energy variation due to the reversal of an isotropic 
domain of aligned spins with linear size $l$ in $D$ dimensions is of the order 
\begin{equation}
\Delta E \sim 2J l^{D-1} - 2l^{\frac{D+1}{2}} Y
\; ,
\label{eq:ImryMa}
\end{equation}
with  $Y$ representing a Gaussian random variable, $Y = \mathcal{N}(0,\sigma^2)$.  
The first term is the energy cost due to the inclusion of a domain wall with length of the 
order of $l$ and the second term is the energy gain that one can achieve from the bulk of the domain 
due to the correlated random fields. 

The excess energy $\Delta E$ in Eq.~(\ref{eq:ImryMa}) is interpreted as a function of $l$. This function has a 
maximum at a given $l$ as long as $D>D_\ell$.
Accordingly, the lower critical dimension in a RFIM with columnar correlated fields is $D_\ell=3$, higher than the one with i.i.d. 
local random fields, which is $D_\ell=2$. Therefore the correlated random fields are 
even more efficient in destroying the ferromagnetic order than the perfectly random ones, as could have been
expected.

Still, this reasoning only applies in the thermodynamic limit. We may still see a pseudo ferromagnetic phase in small systems. For this reason, we will propose that a  pseudo transition survives under the columnar random fields and study it with an effective 1D model before presenting a dynamic analysis that gives support to this assumption.

\subsection{The 1D disordered model}
\label{subsec:1Ddisordered}

We now have an effective 1D Random Field Ising Model (RFIM), with the A columns of size $N$ considered as $N/2$ spins taking values
$N s_j=N (\pm 1)$, leading to an effective coupling constant $\Jeff\propto N$ between the spins $s_j = \pm 1$, 
and i.i.d. random fields with absolute value of order $\sigma= {\mathcal O}(1)$ that couple linearly to the Ising spins.
Its Hamiltonian is
\begin{equation}
\Heff(T) = -\Jeff(T,N) \sum_{j=1}^{N/2} s_j s_{j+1} - N\sum_{j=1}^{N/2} h_j s_j
\; .
\label{eq:1DRFIM}
\end{equation}
For each choice of the $h_j$'s we can compute the partition function, the free energy density and the magnetisation 
and then average over the different realisations of disorder. 

Because at $T=0$, $\Jeff=0$, see Eq.~(\ref{eq:Jeff}), at zero temperature 
the macro-spins are uncoupled and simply align with their associated magnetic field $h_j$.  This single ground state still has magnetisation $M'_{GS} = 0$ because in the infinite size limit, half of the $h_j$'s point up and half down. 
Nonetheless, this ground state has now a lower  
energy than the one of the model without disorder ($E_{\rm GS} = 0$); more precisely, 
\begin{equation}
E'_{\rm GS} = -N\sum_{j=1}^{N/2} \lvert h_j \rvert
\; .
\end{equation}
In the  $N \gg 1$ limit, using $\mathbb{E}[\lvert h_j \rvert] = \sqrt{2/\pi} \ \sigma$ 
and the central limit theorem 
\begin{equation}
E'_{\rm GS} \simeq -\sqrt{1/(2\pi)} \ \sigma \; N^2
\; .
\end{equation}

At very  low temperatures $\Jeff$ is very weak and the system is expected to 
stay in this zero magnetization ground state until a sufficiently high temperature is reached \--- and $\Jeff(T,N)$  is made strong enough \--- for some 
ferromagnetic order to appear despite some of the spins having to be anti-aligned with their magnetic field. 
The energy gain by aligning the $N/2$ macro-spins is $E_F = -(N/2)\Jeff (\Tobd^{\rm ran})$.
We can estimate the crossover temperature $\Tobd^{\rm ran}$ by comparing $E_F$ and $E'_{\rm GS}$, 
leading to
\begin{equation}
\Jeff(\Tobd^{\rm ran},N) \sim \sqrt{2/\pi} \; \sigma N
\label{eq:JeffTobd}
\end{equation} 
with $J_{\rm eff}$ still given by Eq.~(\ref{eq:Jeff}).
Using this equation and setting $\epsilon_1 = 1$, we find that for $\sigma = 0.01$ we should have $\Tobd^{\rm ran} \sim 0.4$ and for  $\sigma = 0.005$, 
$\Tobd^{\rm ran} \sim 0.33$. More generally, $\Tobd^{\rm ran}$ is an increasing function of $\sigma$ that vanishes at $\sigma=0$.

We insist upon the fact that the effective 1D RFIM that  we constructed does not have a 
genuine phase transition, in the same way as the conventional 1D RFIM
does not have one either. Still we can use the random fields to create a sharp crossover at a finite temperature.
Our intuition is that this crossover should be reminiscent of a first order phase transition
because there is no continuity in the two different states of lower energy before and after the crossover. 

We can estimate the length of the system $N_{IM}$ beyond which the pseudo ferromagnetic phase ceases to exist, that is, when flipping a macroscopic domain can lower the energy of the system ($\Delta E < 0$). Thinking in terms of the 1D effective model with the coupling constant  $\Jeff$, the reversal of a domain of 
length $L$ implies an energy cost equal to $4J_{\rm eff}$ and an eventual energy gain equal to $-N\sigma \sqrt{L/(2\pi)}$ due to the random fields. These
two scales are equal for 
\begin{equation}
L_{IM} \sim \bigg(\frac{4\Jeff \sqrt{2\pi}}{N \sigma}\bigg)^2 \; ,
\label{eq:Nc}
\end{equation}
and gives an order of magnitude of the system length, $N_{IM}\simeq L_{IM}$ beyond which ferromagnetic ordering cannot be sustained.


A simple way to compute the equilibrium properties of the 1D effective model 
with periodic boundary conditions and random fields is to use the exact renormalisation 
decimation procedure~\cite{le1999random,dasgupta1980low,IgloiMonthus}. 
Starting from the partition function 
\begin{equation}
Z = \sum_{s_0,s_1,...s_{N/2-1}} \exp \bigg({\sum\limits_{j=0}^{N/2-1}K_j s_j s_{j+1}+ \sum\limits_{j=0}^{N/2-1} H_j s_j}\bigg)
\; , 
\label{eq:Z-disordered}
\end{equation}
where we set $K_j = \beta \Jeff$ and $H_j =\beta h_j$, we can sum over the odd spins and rewrite it in the same form
\begin{equation}
\begin{split}
Z &= \sum_{s_0,s_2,...s_{N/2-2}} \exp\Bigg(\sum_{k=0}^{\frac{N}{4}-1} H_{2k} s_{2k}\Bigg)\prod_{j=0}^{\frac{N}{4}-1} \sum_{s_{2j+1} = \pm 1} \exp(K_{2j}(s_{2j} + s_{2j+2} + H_{2j+1})s_{2j+1})\\
 &= \Bigg(\prod_{k=0}^{\frac{N}{4}-1} c_{2k+1} \Bigg) 
 \sum_{s_0,s_2,...s_{N/2-2}} \exp\Bigg(\sum_{j=0}^{\frac{N}{4}-1}K_{2j}^{'}s_{2j} s_{2j+2} + (H_{2j} + H^{'}_{2j} + H^{'}_{2j+2}) s_j\Bigg)
\end{split}
\label{eq:Z-fact-disordered}
\end{equation}
with $K^{'}$ the rescaled coupling constant and $H^{'}$ the extra magnetic field we add to rescale $H$.
Equating Eq.~(\ref{eq:Z-disordered}) and Eq.~(\ref{eq:Z-fact-disordered}) we find the system of equations
\begin{equation}
\begin{split}
&c_{2j+1} \ e^{K^{'}_{2j} + H^{'}_{2j} + H^{'}_{2j+2}} \ \, = 2\cosh(K_{2j} + K_{2j+1} + H_{2j+1})
\; , \\
&c_{2j+1} \ e^{K^{'}_{2j} - H^{'}_{2j} - H^{'}_{2j+2}} \ \, = 2\cosh( -K_{2j} - K_{2j+1} + H_{2j+1})
\; , \\
&c_{2j+1} \ e^{-K^{'}_{2j} + H^{'}_{2j} - H^{'}_{2j+2}} = 2\cosh( K_{2j} - K_{2j+1} + H_{2j+1})
\; , \\
&c_{2j+1} \ e^{-K^{'}_{2j} - H^{'}_{2j} + H^{'}_{2j+2}} = 2\cosh(- K_{2j} + K_{2j+1} + H_{2j+1})
\; , 
\end{split}
\end{equation}
for $j = 0,...,\frac{N}{4} -1$.
We iterate the decimation until there are only two spins left in the system: $s_0$ and $s_{N/4}$ and we then compute $Z$ for the $4$ configurations of the decimated system. 
From it we derive the free-energy and the mean magnetization
\begin{equation}
F = -\frac{1}{\beta} \ln Z \qquad \text{and} \qquad M = \left. \frac{ \partial F}{\partial (\delta h)} \right|_{{\delta h}=0}
\; ,
\end{equation}
where $\delta h$ is an infinitesimal shift added as a global perturbing magnetic field. 

\begin{figure}[h!]
\begin{center}
\subfloat[]{\includegraphics[scale=0.55]{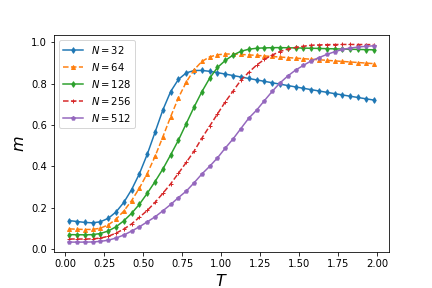}}
\subfloat[]{\includegraphics[scale=0.55]{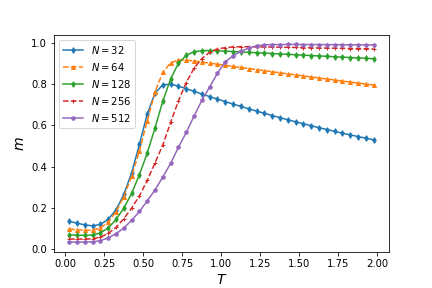}}
\caption{\small Renormalisation group calculation of the mean magnetization density $m = M/N$ 
in the 1D Random Field Ising model (\ref{eq:1DRFIM})
with different system sizes given in the key. (a) $\sigma = 0.01$ and (b) $\sigma = 0.005$
with $\sigma^2$ the variance of the Gaussian distribution from which the magnetic fields are drawn.
}
\label{fig:magn-dens-1D}
\end{center}
\end{figure}

In Fig.~\ref{fig:magn-dens-1D} we observe that, in all cases, the magnetisation density $m$ smoothly increases from $0$ to a value close to $1$. 
For small $N$, $N\leq 128$,  the curves are non-monotonic and $m$ decays again after reaching a maximum. Instead, for sufficiently large $N$, say $N\geq 512$, $m$ monotonically approaches $1$. However, for these large system sizes there is no crossing of curves, of the kind expected in a phase transition. This confirms that the random fields may destroy the 2D ferromagnetic phase in the 
limit $N \rightarrow \infty$, as the Imry-Ma argument that we present in Sec.~\ref{sec:ImryMa} 
shows that indeed occurs. For these sizes, the curves still approach $m=1$ because $\Jeff$ increases with temperature. However, at fixed $T$, the ferromagnetic order is lowered as the size of the system is increased
and one can argue it will disappear in the infinite size limit. If we ignore the fact that there is a strong size dependence in our results, we can still reckon that the magnetisation reaches, say, $0.5$ in a system with 
$N=256$ and $\sigma=0.01$ (a) at $\Tobd^{\rm ran} \sim 0.75$ while it takes the same value in  a system with 
the same system size and $\sigma=0.005$ (b) at a lower temperature, $\Tobd^{\rm ran} \sim 0.6$. This trend is in agreement 
with the estimate in Eq.~(\ref{eq:JeffTobd}), and the numerical values are not too far from the ones given in the text right below this equation.

\subsection{Dynamics}
\label{subsec:dynamics}

In order to confirm the quasi ferromagnetic order reached by the ObD mechanism in a finite range of non-zero temperatures, we focus now on the quench dynamics of the bidimensional model with random columnar magnetic fields, following the evolution of different initial conditions at the target temperatures. 
To study the 2D model we implement a Monte Carlo simulation using the Metropolis algorithm \cite{metropolis1953equation}. 
A time-step is defined as $N^2$ random flip attempts as the system is of size $N\times N$. 
For this simulation, we took the parameters $\Jaa = 2$, $\Jbb = -1$ and $\Jab = 0.75$ 
to keep the energy of the first excitation at $\epsilon_1 = 4(\lvert \Jbb \rvert - \Jab) = 1$, 
and to make the energy between the ground state and the second excited state much larger 
$\epsilon_2 = 4\lvert \Jbb \rvert = 4$. The critical temperature between the ferromagnetic and 
paramagnetic phases of the pure Domino Model, see Eq.~(\ref{eq:critical-temp}), is $T^{\rm pure}_c \simeq 1.40$
for these parameters.

\subsubsection{Quenches from high temperatures}
\label{subsec:highTquench}

Here we investigate  the dynamics following the usual quench protocol~\cite{bray2002theory,Puri09-article,CorberiPoliti}: 
starting from a completely random high temperature initial state, 
$s_{i,j} =\pm 1$ with probability $1/2$, we evolve it with the Metropolis rule at $T=1$, 
where the system should tend to order ferromagnetically for the finite system sizes used here.
Indeed, we estimated the temperature above which no ferromagnetic ordering should be reached to be 
$T^{\rm ran}_c \sim 1.35$ using several runs of the Monte Carlo code for different temperatures and sizes
(not shown). This value is close the one found using Eq.~(\ref{eq:critical-temp}), $T^{\rm pure}_c = 1.4$, for the pure Domino Model 
considering it should be a bit lower in our case as an effect of disorder. Also, using Eq.~(\ref{eq:Nc}), we find that the Imry-Ma length 
is of order $N_{IM}\sim 10^4$, ensuring that we are below this length in the simulations and that 
the system should tend to order ferromagnetically for the sizes accessible in numerical simulations. 
We recall that, for the model with random columnar fields,  
$T^{\rm ran}_{\rm ObD} \simeq 0.4$ for $\sigma = 0.01$ 
and $T^{\rm ran}_{\rm ObD} \simeq  0.33$ for $\sigma = 0.005$. 

\vspace{0.25cm}

\noindent
{\it Snapshots}

\vspace{0.25cm}

\begin{figure}[h!]
\captionsetup[subfigure]{labelformat=empty}
\begin{center}
\subfloat[$t=0$]{\includegraphics[scale=0.5]{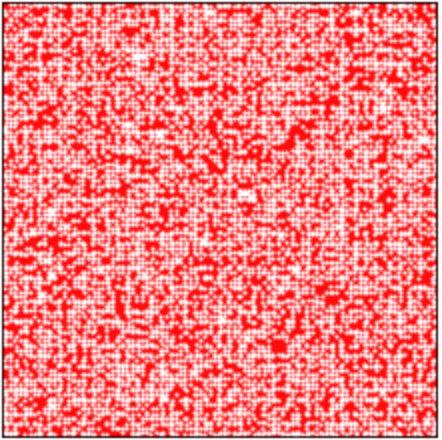}}\hspace{0.1cm}
\subfloat[$t=2^8$]{\includegraphics[scale=0.5]{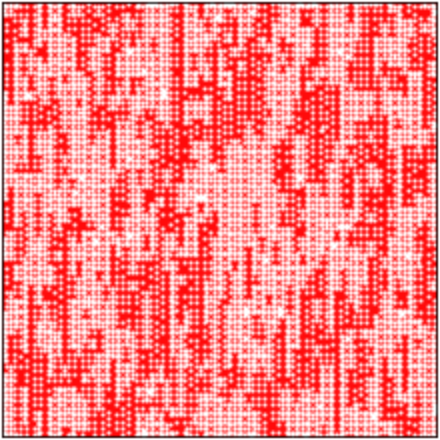}}\hspace{0.1cm}
\subfloat[$t=2^{16}$]{\includegraphics[scale=0.5]{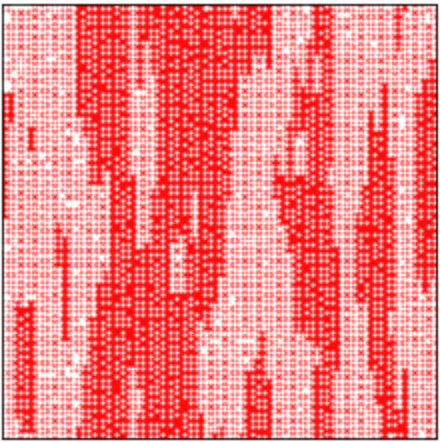}}\hspace{0.1cm}
\subfloat[$t=2^{20}$]{\includegraphics[scale=0.5]{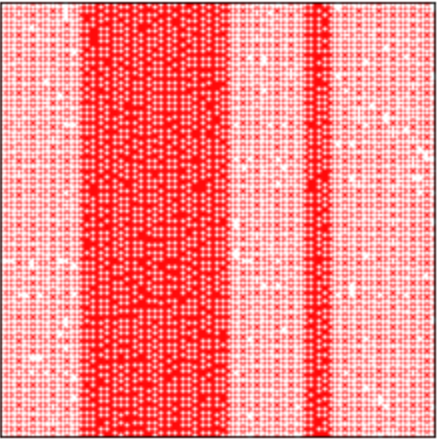}}\hspace{0.1cm}\\
\subfloat[$t=0$]{\includegraphics[scale=0.5]{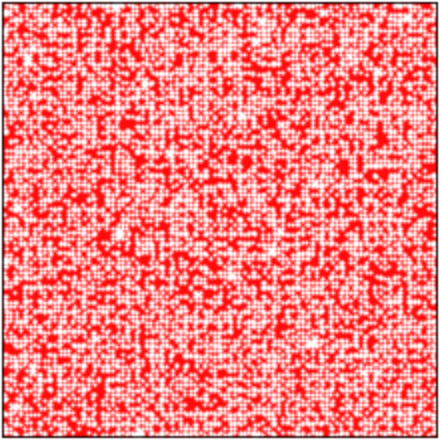}}\hspace{0.1cm}
\subfloat[$t=2^8$]{\includegraphics[scale=0.5]{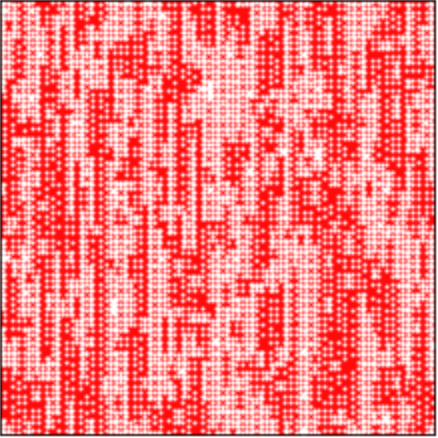}}\hspace{0.1cm}
\subfloat[$t=2^{16}$]{\includegraphics[scale=0.5]{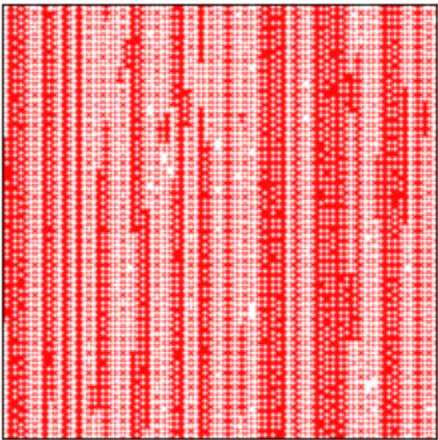}}\hspace{0.1cm}
\subfloat[$t=2^{20}$]{\includegraphics[scale=0.5]{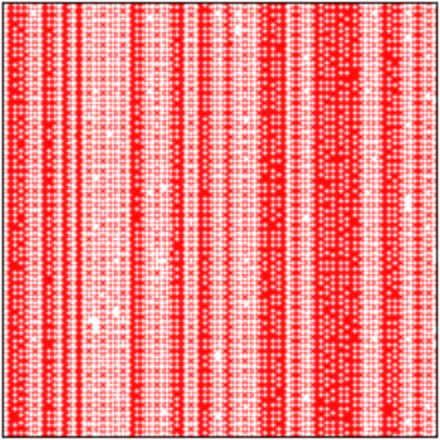}}\hspace{0.1cm}
\caption{\small 
Snapshots of the system with $N=128$ after a quench from a random initial condition across the ferromagnetic 
transition (pseudo in the random problem) to $T=0.35$ which is, moreover, also lower 
than $\Tobd$ in the disordered model. 
The first line shows four representative snapshots of the instantaneous state of the pure model and the second line the same for  the model with quenched random columnar fields with $\sigma = 0.08$ and $T^{\rm ran}_{\rm ObD}\simeq 0.90$.
The time at which the images were stored are indicated below them.
}
\label{fig:snapshots}
\end{center}
\end{figure}

The dynamics of frustrated magnets are expected to be slower than the 
ones of the pure counterparts~\cite{Walter2008,Walter2009}
and in many cases they can also be anisotropic~\cite{Grousson01,Mulet07,Levis12,Levis13,Cannas18,Udagawa18}. 
Indeed, the Domino Model is essentially anisotropic 
and the growth of order should reflect this anisotropy. More precisely, ferromagnetic
ordering along the A columns in the horizontal and vertical directions may, in principle, occur in different time scales, as well as  
anti-ferromagnetic ordering along the B columns. We focus on the growth of ferromagnetic order on A columns.

\begin{figure}
\hspace{1cm}
\subfloat[]{\includegraphics[scale=0.7]{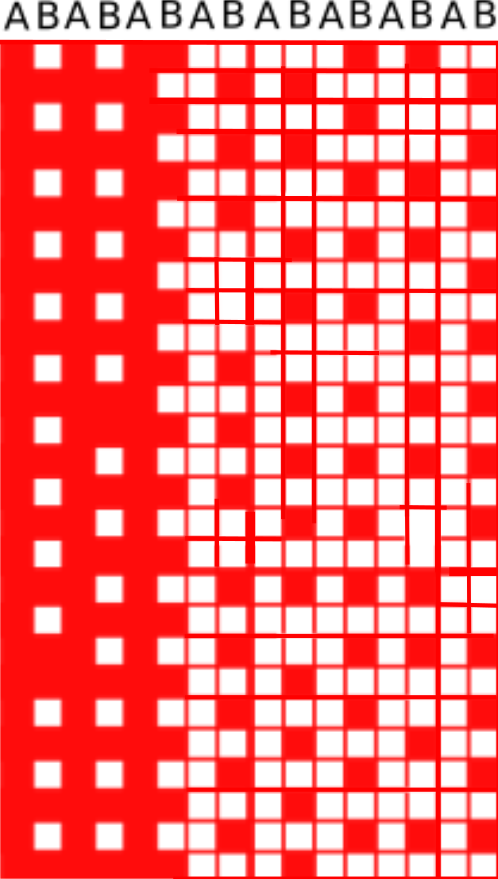}}
\hspace{1.5cm}
\subfloat[]{\includegraphics[scale=0.7]{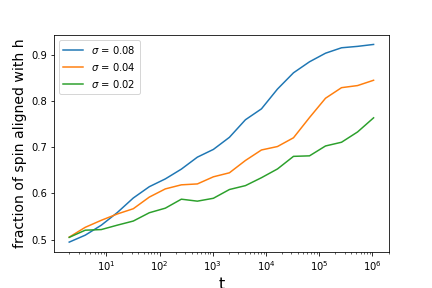}}
\caption{\small (a) Zoom on a snapshot. The different behaviour of A and B columns is clear here.
(b) Fraction of the spins of A columns aligned with the local random magnetic field following the quench 
dynamics of the system with columnar random fields. The upper curve corresponds to the case 
showed in Fig.~\ref{fig:snapshots}.}
\label{fig:zoom}
\end{figure}

Figure~\ref{fig:snapshots}
displays the evolution of a system with $N=128$, quenched from a totally disordered initial 
condition and evolved at $T=0.35$. Red and white cells represent up and down spins.
The first row presents four snapshots 
of the pure Domino Model at the times written below the images ($T=0.35 <T^{\rm pure}_c$ in this case). 
The initial state is fully disordered with as many 
up as down spins placed at random in the box. The system progressively orders and, as it is 
clear from the later images, it does faster in the vertical direction. A typical length of domains in the horizontal 
direction is also growing at a slower speed. Once flat interfaces between the up and down domains 
are created it will take much longer to kill them and fully order the sample ferromagnetically on all A 
columns. More details of the configurations can be seen in the zoom in Fig.~\ref{fig:zoom}(a).

The snapshots of the pure model can be confronted to the ones of the model with the columnar random
fields that are shown in the second row of Fig.~\ref{fig:snapshots} ($T=0.35 <T^{\rm ran}_{\rm ObD}\simeq 0.90$ in this case).
 Globally, the evolution is similar to  the one of the pure model
although some quantitative differences, as the fact that the horizontal extent of the ferromagnetic 
domains is shorter in the random model, are easy to spot. The reason for this is the pinning character 
of the random fields, which is further exhibited in Fig.~\ref{fig:zoom}(b). 

The plot in Fig.~\ref{fig:zoom}(b) shows the evolution of the fraction of spins of A columns that are aligned with their local columnar magnetic field. This fraction increases with time as the system approaches equilibrium and with the typical strength of the fields, $\sigma$. The blue curve is associated with the evolution of the system we follow on the second row of Fig.~\ref{fig:snapshots} and shows that, in the last snapshot ($t=2^{20}$), more than $90\%$ of the spins are already aligned with their magnetic field, probing the pinning (and disordering) character of the latter. 
We deduce that the spin domains on this snapshots are mostly due to parts of the system where the $h_j$ have the same sign.

The figures suggest that while the horizontal length scale of the domains between flat walls is of the order of 
the system size in the pure model,   the domains are of finite horizontal size in the disordered case.

\vspace{0.25cm}

\noindent
{\it Magnetisation and correlations}

\vspace{0.25cm}

\begin{figure}[h!]
\begin{center}
\includegraphics[scale=0.6]{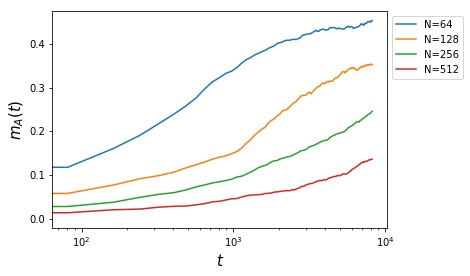}
\caption{\small Monte Carlo dynamics at $T_{\rm ObD} < T=1 < T^{\rm ran}_c$ of the 2D Domino Model with columnar random fields with 
$\sigma=0.01$.  Time evolution of the mean magnetisation density of the A columns, $m_A(t)$ defined in 
Eq.~(\ref{eq:magn-A}), after quenches from a fully random initial condition
across the PM-FM crossover at $T^{\rm ran}_c$ and above the one at $T^{\rm ran}_{\rm ObD}$. Different curves
correspond to different sizes given in the key.
}
\label{M-evol-PMFM}
\end{center}
\end{figure}

After the generic discussion of the snapshots in Fig.~\ref{fig:snapshots},
in Fig.~\ref{M-evol-PMFM} we show the time evolution of $m_A$ defined as 
\begin{equation}
m_A(t) = \frac{2}{N^2}\Big\langle\Big\lvert\sum_{k_A=1}^{N^2/2} s_{k_A}(t) \Big\rvert\Big\rangle
\label{eq:magn-A}
\end{equation}
with $\langle \dots \rangle$ the average over many realisations of the dynamics and $k_A$ running over the A spins indices only.
For  $N=512$ the magnetisation remains smaller than 0.1 until $t \simeq 10^4$. The analysis of the coarsening
process will be done for such linear system size ensuring that the evolution remains sufficiently far from any possible 
equilibration.

\begin{figure}[t!]
\begin{center}
\subfloat[]{\includegraphics[scale=0.55]{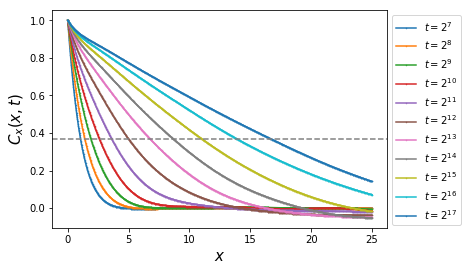}}
\subfloat[]{\includegraphics[scale=0.55]{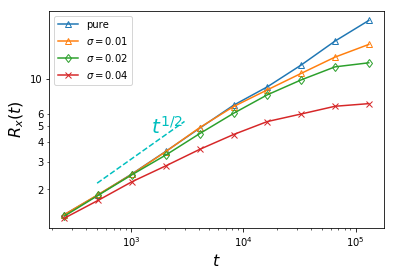}}
\caption{\small 
Monte Carlo dynamics  after a quench (cooling) from a completely random initial condition across the 
PM-FM and above the Order by Disorder crossovers to $T=1$.
Square system with linear length $N=512$.
(a) Horizontal correlation function $C_x(x,t)$ as a function of $x$ for a system with columnar random fields ($\sigma=0.01$), at different times given in the key.
(b) Evolution of the typical growing correlation length $R_x(t)$ for systems with different standard variation of the 
random fields given in the key.
}
\label{fig:corr-PMFM}
\end{center}
\end{figure}

The plot in Fig.~\ref{fig:corr-PMFM}(a) is representative of the coarsening dynamics across a second order phase transition~\cite{bray2002theory,Puri09-article,CorberiPoliti}. We display the horizontal correlation function 
of the spins sitting on the A columns
\begin{equation}
\begin{split}
C_x(x,t) &=\frac{ \frac{2}{N^2} \Big( \sum_{i,j} s_{2i,j} s_{2(i+x),j} - \big( \sum_{i,j} s_{2i,j} \big)^2 \Big)}
{1 - \big( \sum_{i,j} s_{2i,j} \big)^2}
= \frac{\frac{2}{N^2} \sum_{i,j} s_{2i,j} s_{2(i+x),j} - m_A(t)^2}{1-m_A(t)^2}
\end{split}
\end{equation}
of a system with $N=512$ for which the ferromagnetic magnetisation density of these columns at the longest
time $t\simeq 10^5$ should be of order of $m_A \simeq 0.1$, see Fig.~\ref{M-evol-PMFM}. 
The system progressively orders, and this is represented by a $C_x(x,t)$ that decays  to $0$ with distance 
in a slower manner for increasing times. These curves can be compared, for example, to the ones in 
Fig. 17 in Ref.~\cite{sicilia2007domain}, where similar data for the 2D Ising Model are 
shown.
We obtain the typical growing correlation length in the $x$-direction from the standard criterion $C_x(R_x(t),t)\sim1/e$ 
(see the horizontal  dotted line in Fig.~\ref{fig:corr-PMFM}(a)). We then plot the evolution of $R_x(t)$ with time in panel (b). 
We find that at short time scales the pure and disordered Domino Models have 
$R_x(t) \propto t^{1/2}$, as expected for the curvature driven dynamics of a non-conserved scalar order parameter system. 
The various curves correspond to different strengths of the random fields, as quantified by their standard deviation
$\sigma$ given in the key. At the longest time scales that we show the growth in the model with random fields saturates, to a value that 
decreases with increasing $\sigma$. The annihilation of these domain walls should involve much longer time scales (see, for 
example~\cite{Redner02,Blanchard17}, for their study in the pure 2D Ising model) and it needs thermal activation to create a 
bump on the otherwise flat interfaces that, moreover, are pinned by the random fields.

\begin{figure}[H]
\hspace{-0.5cm}
\subfloat[]{\includegraphics[scale=0.55]{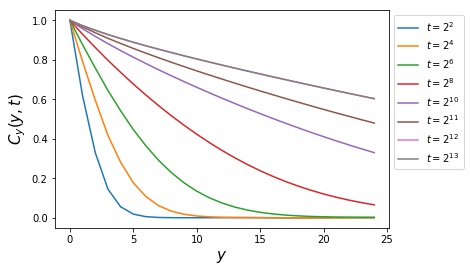}}
\subfloat[]{\includegraphics[scale=0.55]{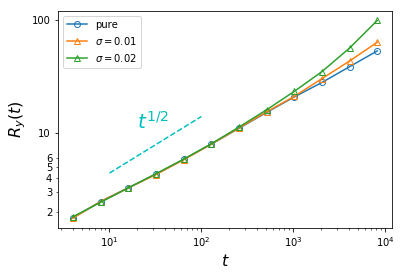}}
\caption{\small Monte Carlo dynamics of the 2D square Domino Model, of linear length $N=512$, 
under columnar random fields with $\sigma=0.01$. 
Dynamics after a quench from a completely random initial condition across the PM-FM,
and above the Order by Disorder, crossovers to $T=1$. 
(a) Vertical correlation function on A columns, $C_{y}(y,t)$, as a function of $y$ for different times given in the key.  
(b) The growing correlation length, $R_y(t)$,  in the vertical direction. 
}
\label{vertical_corr_PMFM}
\end{figure}

Figure~\ref{vertical_corr_PMFM} confirms that the system orders faster vertically than horizontally. In panel (a) we present the vertical 
correlation function of the spins belonging to columns  A
\begin{equation}
C_y(y,t) = \frac{\frac{2}{N^2} \sum_{i,j} s_{2i,j} s_{2i,j+y} - m_A(t)^2}{1-m_A(t)^2}
\end{equation}
while in panel (b) we represent the corresponding growing length as a function of time.

\subsubsection{Heating from the disordered ground state}
\label{subsec:ObDquench}

\begin{figure}[h!]
\captionsetup[subfigure]{labelformat=empty}
\begin{center}
\subfloat[$t=0$]{\includegraphics[scale=0.5]{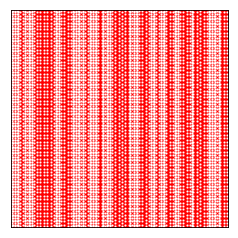}}
\subfloat[$t=2^8$]{\includegraphics[scale=0.5]{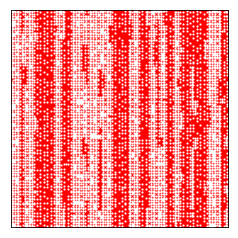}}
\subfloat[$t=2^{16}$]{\includegraphics[scale=0.5]{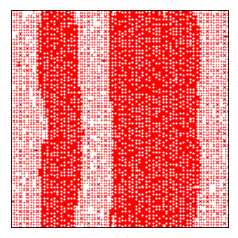}}
\subfloat[$t=2^{22}$]{\includegraphics[scale=0.5]{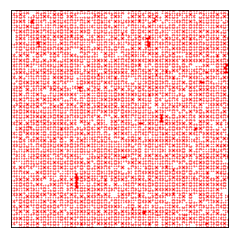}}\\
\subfloat[$t=0$]{\includegraphics[scale=0.5]{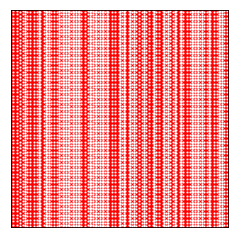}}
\subfloat[$t=2^8$]{\includegraphics[scale=0.5]{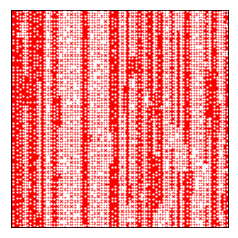}}
\subfloat[$t=2^{16}$]{\includegraphics[scale=0.5]{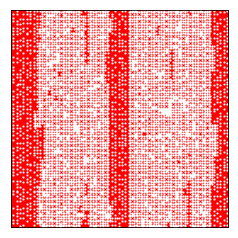}}
\subfloat[$t=2^{22}$]{\includegraphics[scale=0.5]{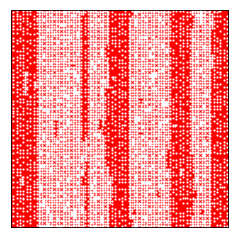}}
\caption{\small Snapshots of a system of size $N=128$ after a sudden increase in temperature from the disordered ground state
to $T=1$. In the pure model, this temperature is below $T^{\rm pure}_c$. In the disordered one ($\sigma=0.01$), 
it is in between the pseudo critical 
temperatures $\Tobd^{\rm ran}$ and $T_c^{\rm ran}$. 
The time at which the images were stored are indicated below them.}
\label{fig:snapshots-heat}
\end{center}
\end{figure}

We now investigate the dynamics across the ObD crossover itself, starting from the ground state (corresponding to $T=0$) 
and fixing the working temperature to $T=1$ as in the sub-critical quenches discussed in Sec.~\ref{subsec:highTquench} where the system, for the sizes we use, should 
eventually approach a ferromagnetic configuration. Snapshots in Fig.~\ref{fig:snapshots-heat} show an example of these dynamics.
In the pure system (top panels) the final configuration is one in which the system ordered ferromagnetically on the A columns with 
-1 spins. In the disordered case (bottom panels) the dynamics is slower and the stationary state has not been reached yet.
Data for the pure model are gathered using, 
for each Monte Carlo run, an initial state chosen randomly among the collection of all  possible ones. Instead, the 
simulations with random fields are started from the unique ground state, which is determined for each run by the magnetic fields that
we draw from the Gaussian distribution. The average is then computed over random fields and/or Monte Carlo random numbers.

\begin{figure}[H]
\begin{center}
\subfloat[]{\includegraphics[scale=0.55]{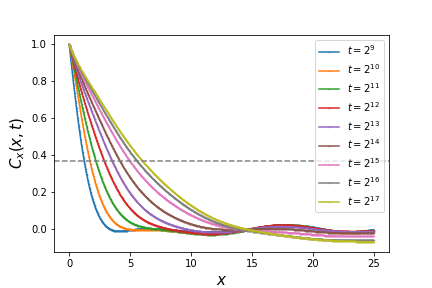}}
\subfloat[]{\includegraphics[scale=0.55]{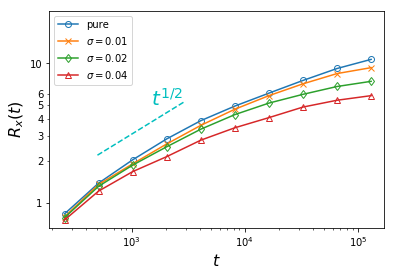}}
\caption{\small Heating across the ObD crossover in a square 2D system with linear length $N=512$, 
and columnar random fields with $\sigma=0.01$,
evolving from one of the zero temperature ground states.
(a) Horizontal correlation  $C_x(x,t)$ as a function of $x$, at $T=1$, $\sigma=0.01$, 
and for different times given in the key. 
(b) Evolution of the typical growing correlation length $R_x(t)$ for different values of the disorder strength. 
}
\label{fig:corr-ObD}
\end{center}
\end{figure}

In Fig.~\ref{fig:corr-ObD} one finds the horizontal correlation functions as functions of distance, for different times, in panel (a). 
The curves have the same qualitative behaviour as the ones already shown for the quenches from the infinite 
temperature state. 
However, the growing length is pretty different from, and much slower than, 
the usual $t^{1/2}$ curvature driven form, as can be seen in panel (b).

 \section{Staggered columnar magnetic fields}
\label{sec:staggered}

In order to have a phase transition towards a ferromagnetically order state upon increasing  temperature, 
circumventing the Imry-Ma argument, 
we no longer use random fields, but alternate columnar magnetic fields $h_j = (-1)^j h$. 
Because the $h_j$ are not random but staggered, the formation of macroscopic 
reversed ferromagnetic domains is no longer possible. The drawback is that we lose some specificity of the 
ObD phenomenon because the zero-temperature ground state is now antiferromagnetic as the staggered magnetic fields impose.  
Still, the strategy is to use these fields as a probe to exhibit the underlying conventional ObD transition. 
The idea is 
to impose an antiferromagnetic equilibrium state at very low temperature, that would be replaced by the 
ferromagnetic one at a first order phase transition taking place at a finite temperature 
below the one at which the system reaches the paramagnetic high temperature phase. 

\subsection{The 1D model}

The Hamiltonian of the effective 1D  model under staggered local fields is
\begin{equation}
\Heff(T) = -\Jeff(T,N) \sum_{j=1}^{N/2} s_j s_{j+1} - Nh\sum_{j=1}^{N/2} (-1)^j s_j
\end{equation}
with $\Jeff(T,N)\propto N$, as given in Eq.~(\ref{eq:Jeff}).

\vspace{0.2cm}

\begin{figure}[H]
\begin{center}
\includegraphics[scale=0.65]{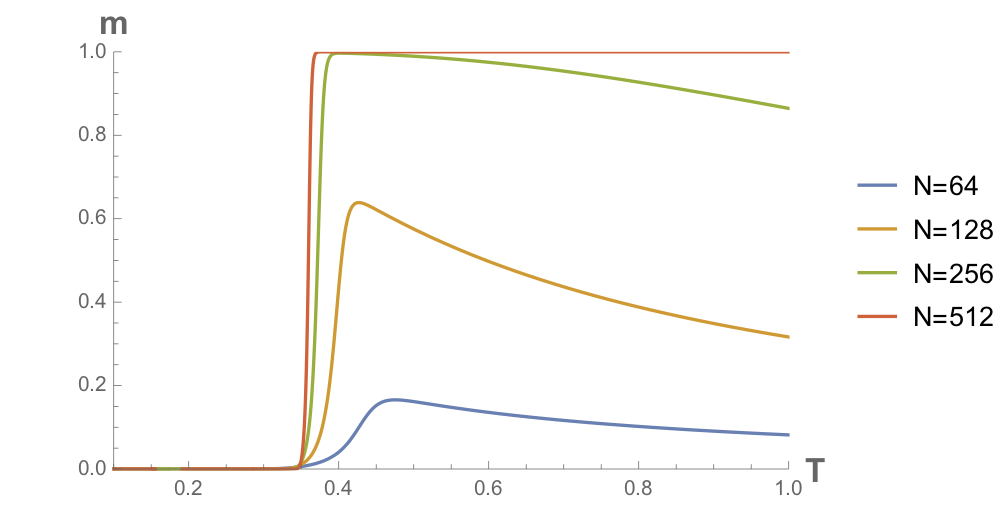}
\caption{\small Mean magnetization of the 1D model with staggered magnetic fields
with amplitude $h=0.01$,
for different system sizes given in the key. 
}
\label{fig:magn-stagg}
\end{center}
\end{figure}

We compute the mean magnetisation using the transfer matrix method with a matrix $\mathcal{T}=W_1 W_2$ representing a block of two columns with $W_1$ for a column with a positive magnetic field and $W_2$ for a negative one
   \begin{align*}
& W_1 = 
   \begin{pmatrix}
e^{\frac{N}{4}\ln(1+e^{-1/T}) + \frac{N\delta h}{T}} & e^{-\frac{N}{4} \ln(1+e^{-1/T}) + \frac{Nh}{T}} \\
e^{-\frac{N}{4} \ln(1+e^{-1/T}) - \frac{Nh}{T}}  & e^{\frac{N}{4}\ln(1+e^{-1/T}) - \frac{N\delta h}{T}} 
   \end{pmatrix} 
   \; , 
   \\
   & W_2 = \begin{pmatrix}
   e^{\frac{N}{4}\ln(1+e^{-1/T}) + \frac{N\delta h}{T}} & e^{-\frac{N}{4} \ln(1+e^{-1/T}) - \frac{Nh}{T}} \\
e^{-\frac{N}{4} \ln(1+e^{-1/T}) + \frac{Nh}{T}}  & e^{\frac{N}{4}\ln(1+e^{-1/T}) - \frac{N\delta h}{T}} 
   \end{pmatrix}
   \; , 
  \end{align*}
with $\delta h>0$ an infinitesimal magnetic field we add to compute the magnetisation.
Writing $\lambda_+$ and $\lambda_-$ the eigenvalues of $\mathcal{T}$, the free energy per spin is 
\begin{equation}
f = -\frac{2T}{N^2} \; \ln\Big( \lambda_+^{N/4} + \lambda_-^{N/4}\Big)
\end{equation}
and the mean magnetization per spin $m$ is
\begin{equation}
m = - \frac{\partial f}{\partial(\delta h)}\bigg|_{\delta h=0} \; .
\end{equation}

We find a transition temperature $\Tobd^{\rm col} = 0.35$ with $h=0.01$ (see Fig.~\ref{fig:magn-stagg})
which corresponds to what we expect by comparing the interaction energy governed by the effective coupling constant at $\Tobd^{\rm col} $ 
and the energetic contribution of the magnetic field 
\begin{equation}
\frac{\Jeff(\Tobd^{\rm col} , N)}{N} = \frac{h}{2} \; .
\end{equation}
We note that, apart from numerical constants, this is the same equation as (\ref{eq:JeffTobd}), 
where $\sigma$ has been replaced by $h$. $\Tobd^{\rm col}$ for the columnar field model is also an 
increasing function of $h$ departing from 0. The numerical data displayed 
in Fig.~\ref{fig:magn-stagg}, which represent $m$ as a function of $T$, confirm the fact that the system
undergoes a first order phase transition at $\Tobd^{\rm col}$.

\subsection{Dynamics}
\label{subsec:dynamics-alternate}

We now turn to the analysis of the quench dynamics of the 2D model with alternate columnar
magnetic fields.

\begin{figure}[h!]
\captionsetup[subfigure]{labelformat=empty}
\begin{center}
\subfloat[$t=0$]{\includegraphics[scale=0.5]{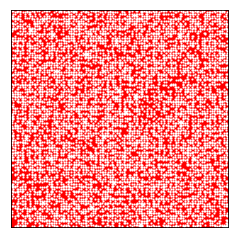}}
\subfloat[$t=2^6$]{\includegraphics[scale=0.5]{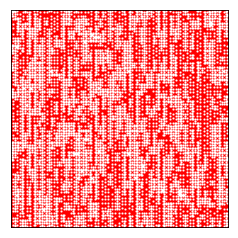}}
\subfloat[$t=2^{12}$]{\includegraphics[scale=0.5]{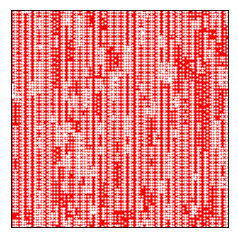}}
\subfloat[$t=2^{18}$]{\includegraphics[scale=0.5]{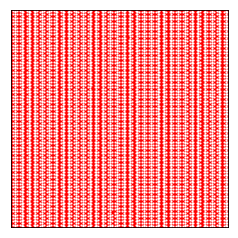}}\\
\subfloat[$t=0$]{\includegraphics[scale=0.5]{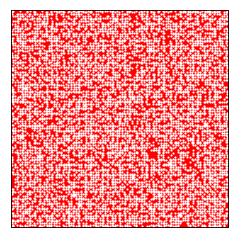}}
\subfloat[$t=2^{6}$]{\includegraphics[scale=0.5]{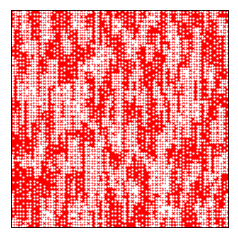}}
\subfloat[$t=2^{12}$]{\includegraphics[scale=0.5]{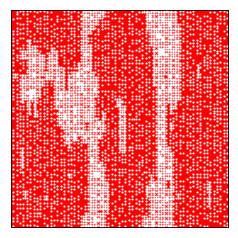}}
\subfloat[$t=2^{18}$]{\includegraphics[scale=0.5]{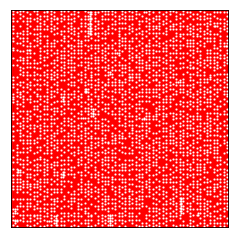}}
\caption{\small Snapshots of the system with $N=128$ and $h=0.08$ after a quench from a random initial condition 
to $T=0.35<\Tobd^{\rm col}$ (first line) and $\Tobd^{\rm col}<T=1<T_c^{\rm col}$ (second line). The time at which the images were stored are indicated below them. 
On the latest image of the first line, the configuration is one of the  ground states with alternate ordering of A columns, whereas on the latest image of the second line, the ordering of A columns is ferromagnetic.
}
\label{fig:snapshots-stagg}
\end{center}
\end{figure}

Figure \ref{fig:snapshots-stagg} shows the evolution of a system with $N=128$ and staggered columnar magnetic fields of strength $h=0.08$, quenched from a disordered initial condition and evolved at $T=0.35<\Tobd^{\rm col}$. If we compare these snapshots with those on Fig.~\ref{fig:snapshots}, we see that as for the two systems studied before (the pure Domino Model and the one with random magnetic fields), the system progressively orders in the vertical direction. The difference here is that the domains are not growing in the horizontal direction because of the pinning character of the alternate magnetic fields. On the last snapshot, we can see that the system reached its equilibrium state at $T=0.35$ which is also a ground state of the pure Domino Model at $T=0$.

\begin{figure}[H]
\begin{center}
\subfloat[]{\includegraphics[scale=0.55]{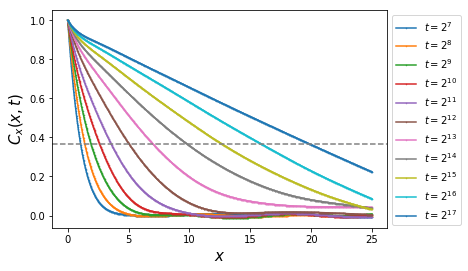}}
\subfloat[]{\includegraphics[scale=0.55]{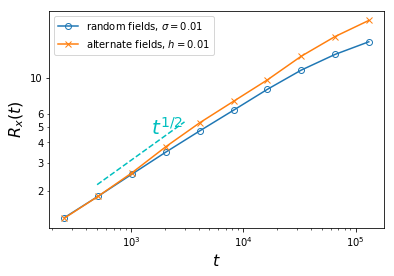}}
\caption{\small Monte Carlo dynamics of the 2D Domino Model with columnar alternate fields ($h=0.01$). 
Dynamics after a quench from a completely random initial condition across the PM-FM transition. 
The evolution is followed at $T=1$, that is, $T_{\rm ObD} < T < T_c^{\rm col}$. 
(a) Horizontal correlation  $C_x(x,t)$ as a function of $x$ for different times given in the key. 
(b) Evolution of the typical growing correlation length $R_x(t)$. 
Square system of linear length $N=512$. 
}
\label{fig:corr-stagg-ObD}
\end{center}
\end{figure}

In Fig.~\ref{fig:corr-stagg-ObD} we display the horizontal correlation functions of the Domino Model with columnar alternate 
fields of strength $h=0.01$ for increasing times given in the key of (a). In panel (b) the growing correlation is reported 
and compared to the $t^{1/2}$ law as well as to the growing correlation of the random fields case with $\sigma = 0.01$.
The data confirm that the system orders ferromagnetically on the A columns in between $\Tobd^{\rm col}$ and $T_c^{\rm col}$.

\clearpage

\section{Conclusion}

The goal of our work was to find a way to displace the thermal ObD transition from 
zero to a non-vanishing temperature. The idea was to  thus render the experimental observation 
of this phenomenon easier. 
To reach this aim we followed two routes, using the Domino Model as the testing ground.

On the one hand, we added well-tuned quenched columnar random fields. These  fields lift the degeneracy of the ground states, 
selecting one that still has zero magnetisation but lower energy than the one under no fields. Consequently, the system is stuck in this state until the temperature is high enough for it to access the large number of first excited states. In this case, the ObD crossover happens at a finite temperature $\Tobd > 0$ but long-range order is suppressed by this type of disorder in the thermodynamic limit. Still, we observed an ObD crossover at a finite temperature for small system sizes using various numerical and theoretical methods that were in good agreement with our predictions. 

In the second approach, we used alternate columnar magnetic fields that do indeed 
 displace the transition at finite temperature. In this case we computed the theoretical ObD 
transition temperature using the transfer matrix method and we confirmed it with dynamic measurements.
We also mentioned some indications that the ObD transition is first order. 

Even though both random and alternate fields impose the ground state, the finite temperature crossover or transition can be used to 
probe the 
ObD phenomenon in the model without applied fields. 
Since the crossover or transition temperatures can be tuned at will, our procedure allows one to probe the ObD mechanism without going to too low temperature, where other kind of energetic contribution might interfere with it.
In conclusion, we think that these methods should be useful to check whether a system exhibits 
the ObD transition. 
  
  \vspace{0.5cm}
  \noindent
  {\bf Acknowledgements}
  We are grateful to P. Guruciaga, J. Restrepo and A. Tartaglia for early discussions of this problem.
  
\bibliographystyle{ieeetr}
\bibliography{References}

\end{document}